\documentclass[aps,prl,superbib,superscriptaddress,amsfonts,graphicx,preprint]{revtex4-1}

\usepackage[utf8]{inputenc}
\usepackage[T1]{fontenc}
\usepackage{tgtermes}
\usepackage{indentfirst}
\usepackage{amsmath, amssymb}
\usepackage{mathtools}
\usepackage{breqn}
\usepackage{graphicx}
\usepackage{float}
\usepackage{bm}
\usepackage{physics}
\usepackage{xcolor}
\usepackage{verbatim}

\usepackage{hyperref}
\usepackage{multirow,xcolor}
\hypersetup{
	colorlinks = true,
	linkcolor = blue,
	linkbordercolor = white,
	citecolor=blue,
}
\usepackage{subfig,caption}
\newcommand{\up}{\uparrow}
\newcommand{\down}{\downarrow}

\makeatletter
\let\cat@comma@active\@empty
\makeatother

\begin{document}
\title{Near Total Electronic Spin Separation as Caused by Nuclear Dynamics: Perturbing a Real-Valued Conical Intersection with Complex-Valued Spin-Orbit Coupling}

\author{Yanze Wu}
\author{Joseph E. Subotnik}
\affiliation{Department of Chemistry, University of Pennsylvania, Philadelphia, Pennsylvania 19104, USA}
\date{\today}

\begin{abstract}
We investigate the nuclear dynamics near a real-valued conical intersection that is perturbed by a complex-valued spin-orbit coupling. For a model Hamiltonian with two outgoing channels, we find that even a small spin-orbit coupling can dramatically affect the pathway selection on account of Berry force, leading to extremely large spin selectivity (sometime as large as 100\%).
Thus, this Letter opens the door for organic chemists to start designing spintronic devices that use nuclear motion and conical intersections (combined with standard spin-orbit coupling) in order to achieve spin selection. Vice versa, for physical chemists, this Letter also emphasizes that future semiclassical simulations of intersystem crossing (which have heretofore ignored Berry force) should be corrected to account for the spin polarization that inevitably arises when dynamics pass near conical intersections.
\end{abstract}

\maketitle

\section{Introduction}
Electronic spin is one of the most fundamental observables in quantum mechanics, and manipulating spin (so-called ``spintronics'') is an enormous and exciting field of research today \cite{Zutic2004,Rocha2005,Fert2008,Chumak2015,Linder2015,Jungwirth2016,Baltz2018}. Even though the energy associated with flipping a single spin is incredibly small (6 $\mu$eV in the presence of a 0.1 T magnetic field), non-trivial spin manipulation can be achieved today through various techniques that center around the coupling between electronic motion and spin dynamics.
Today, there are many physicists studying how giant and tunnel magnetoresistance \cite{Baibich1988,Moodera1995}, spin-orbit torques \cite{Gambardella2011,Brataas2012, Manchon2019}, spin-transfer torques \cite{Berger1996,Slonczewski1996,Ralph2008,Mahfouzi2012,Bajpai2019}, and spin-Hall effects \cite{Hirsch1999,Nikolic2005,Sinova2015} can produce spin polarization either in the presence of external magnetic fields or carefully chosen solid-state materials with some degree of ferromagnetism or both. 
Interestingly, however, recent chiral-induced spin-selectivity (CISS) experiments by Naaman, Waldeck and co-workers \cite{Naaman2012,Naaman2015,Naaman2020} have demonstrated that unusually large electronic spin polarization can also arise when current is passed through chiral molecules without ferromagnetic materials or magnetic fields (and despite very small spin-orbit coupling (SOC) matrix elements). It would appear that, as a community, we still have a great deal to learn about the subtle means by which non-trivial spin effects emerge in practice.    

In this Letter, our goal is to study a different means of achieving spin polarization that, to our knowledge, has not yet been fully explored theoretically (and one that appears to have large experimental consequences). Our focus here will be on the coupling of nuclear motion to electronic spin in the presence of spin-orbit coupling (but without any magnetic fields).
In a recent article, we demonstrated that the interaction of nuclear dynamics with spin orbit coupling can lead to non-zero electronic spin polarization through due to the presence of a state-specific Berry force, and we hypothesized that such effects could be relevant for CISS effects \cite{wu2020chemical}. In particular, we have shown that, for a two-state Hamiltonian of the form (where $r,\theta$ are polar coordinates of nuclear position),
\begin{align} \label{eq:hprev}
    H_{\text{tot}} &= \frac{p^2}{2M} + H_{\up\up} \\
    H_{\up\up} &= \begin{bmatrix}E_A(r,\theta) & V(r,\theta)e^{iWr} \\ V(r,\theta)e^{-iWr} & E_B(r,\theta) \end{bmatrix}
\end{align}
nuclear dynamics on diabat A can lead to spin polarization on diabat B provided that ($i$) there is no spatial inversion between diabats A and B, ($ii$) the nuclei are not in thermal equilibrium on diabat A; ($iii$) the diabatic coupling does not have a constant phase, i.e. $W \ne 0$. In such a case, Ref. \cite{wu2020chemical} demonstrates that the relative difference in the state-to-state transmission rate between up and down electronic spin can be as large as 10\% for a model system with reasonable parameters.
Note that $H_{\down\down} = H^*_{\up\up}$ because of time reversibility \cite{Gould1984}.

For the simple Hamiltonian in Eq.~\eqref{eq:hprev}, the underlying physical mechanism behind any possible
spin polarization here is the Lorentz-like force ($F_B$) arising from the Berry curvature of the electronic surfaces \cite{MichaelVictorBerry1984,Takatsuka2011,Berry1993}. Let $\ket{0}$ and $\ket{1}$ be the ground and excited electronic states of $H_{\up\up}$.
For a nuclear wavepacket moving along $\ket{0}$ with electronic spin up, the Berry force is equal and opposite to what a nuclear wavepacket with an electronic spin down in state $\ket{0}^*$ would feel moving along $H_{\down\down}$\cite{wu2020chemical}:
\begin{align} \label{eq:berry}
    \bm{F}_B^{\up} &= \frac{2\hbar}{M}\Im{\bm{d}_{01} (\bm{p}\cdot \bm{d}_{10})} = \frac{\hbar W}{M}\zeta(r,\theta) [p_y,-p_x] = -\bm{F}_B^{\down} \\
    \zeta(r,\theta) &= \frac{1}{r}\pdv{}{\theta}\Big(\frac{E_A-E_B}{\sqrt{{(E_A-E_B)}^2+4V^2}}\Big)
\end{align}
Here, $\bm{p}$ is the nuclear momentum, $M$ is the nuclear mass, and $\bm{d}_{01} = \bm{\nabla}H_{01}/(E_1 - E_0)$ is the derivative coupling between the two adiabats. 
As a consequence of Berry's magnetic force, a nuclear wavepacket with one spin 
can follow an entirely different trajectory from the identical nuclear wavepacket with the opposite spin. According to Eq.~\eqref{eq:berry}, all spin separation will be proportional to how the diabatic coupling changes phase as a function of position (i.e. the parameter $W$).

Now, a realistic {\em ab initio} electronic Hamiltonian is usually chosen to be of the form $H = H_0 + H_{\text{SOC}}$, where $H_0$ is the real-valued, standard electronic Hamiltonian, and all potentially complex-valued spin-orbit effects are contained in $H_{\text{SOC}}$.
In such a case, if one wishes to map to the reduced Hamiltonian in Eq.~\eqref{eq:hprev}, one can roughly estimate the parameter $W$ from the ratio of the SOC strength to diabatic coupling strength. In other words, $W = \pdv{}{r}\arg{\frac{\abs{\mel{0}{H_{\text{SOC}}}{1}}}{\abs{\mel{0}{H_0}{1}}}}$, where $H_0$ and $H_{\text{SOC}}$ are the (real-valued) spin-free Hamiltonian and the (potentially complex-valued) SOC Hamiltonian, respectively.
Thus, for molecules or molecular assemblies, given that the SOC strengths are usually small (< 1 meV) and diabatic couplings are usually much larger (> 10 meV), one would not expect that $W$ should be very large. And,
as just mentioned, for a reasonably sized $W$, we do not expect more than 10\% polarization (if at all).


With this background in mind, the question we will address in the present Letter is how the argument and the results above change in the presence of a conical intersection (CI) \cite{Yarkony1996} .
This is a very natural question to ask since, in the vicinity of a CI, the spin-free diabatic coupling goes to zero and the derivative coupling is complex-valued and diverges to infinity \cite{Matsika2001,Matsika2001a,Matsika2002,Matsika2002a}. Moreover, CIs are known today to be essential for mediating a vast array of photochemical relaxation processes.
And yet, 
in his seminal paper on geometric magnetism \cite{Berry1993}, Berry argued that the presence of a CI should not lead to drastically different nuclear dynamics:
``These classical effects will however be weak, since the monopole strength is $\pm\frac{1}{2}\hbar$, which vanishes in the classical limit. Moreover, the breakdown of the adiabatic approximation will be greatest at the degeneracies, because of transitions between adiabatic states.'' \cite{Berry1993}

For this reason, in the present manuscript, we will investigate an ``avoided'' complex-valued conical intersection. More precisely, we will investigate a real-valued spin-free Hamiltonian $(H_0)$ for which we find a CI; however, we add to this Hamiltonian a second, constant complex-valued Hamiltonian $(H_{\text{SOC}})$ which formally eliminates the conical intersection.
In this case, we will find that the Berry magnetic force can yield a truly enormous effect,
with spin-separation efficiencies close to 100\% at certain energies.
Furthermore, because the degree of spin polarization depends on the ratio between the SOC and the non-SOC diabatic coupling, and the diabatic coupling vanishes at a CI, we find that a huge amount of spin-polarization can occur even with a very weak SOC.

\section{Model Hamiltonian}
\begin{figure}[H]
    \begin{center}
    \subfloat{\includegraphics[width=0.42\columnwidth]{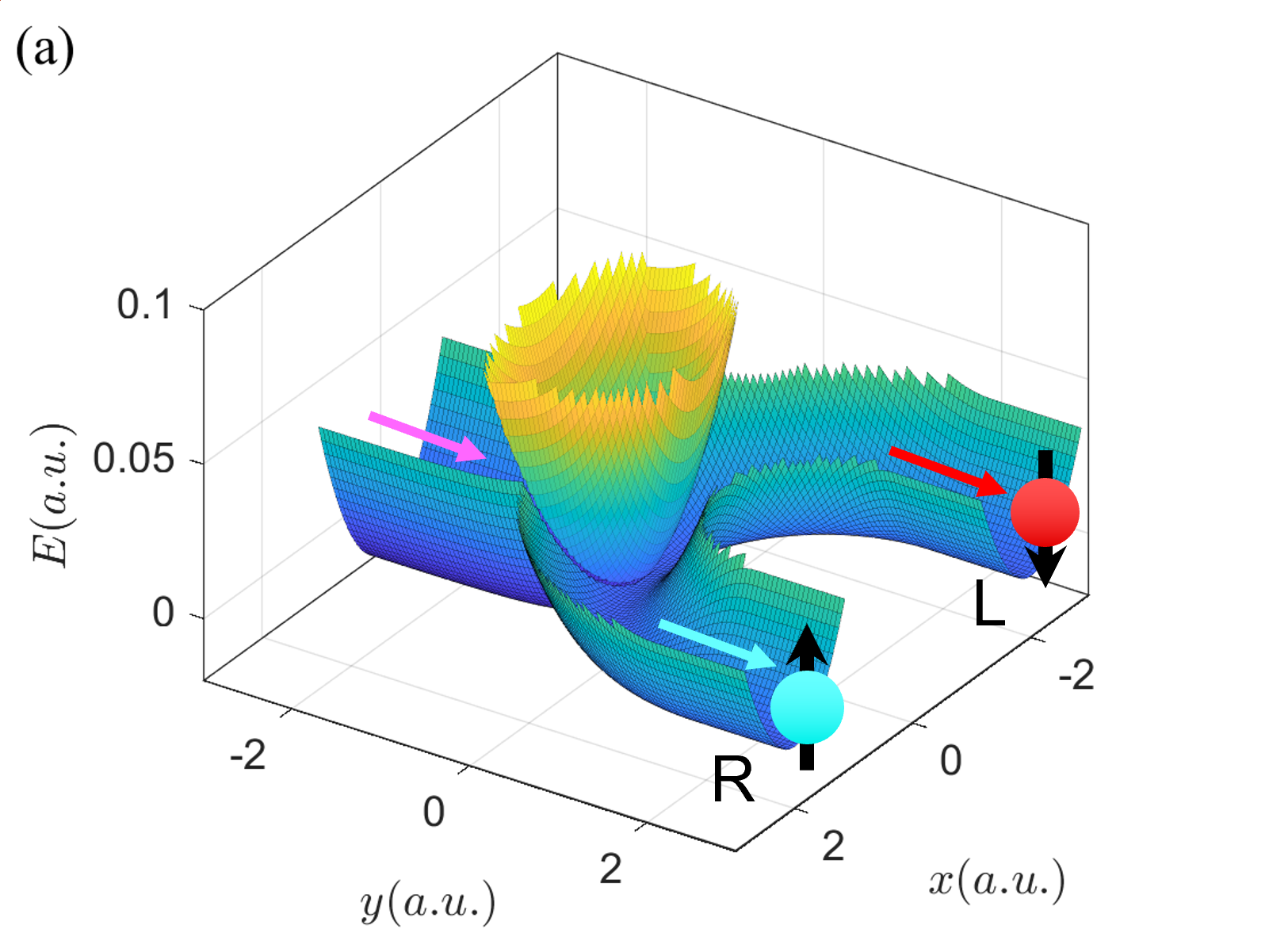}\label{fig:pe}}
    \subfloat{\includegraphics[width=0.42\columnwidth]{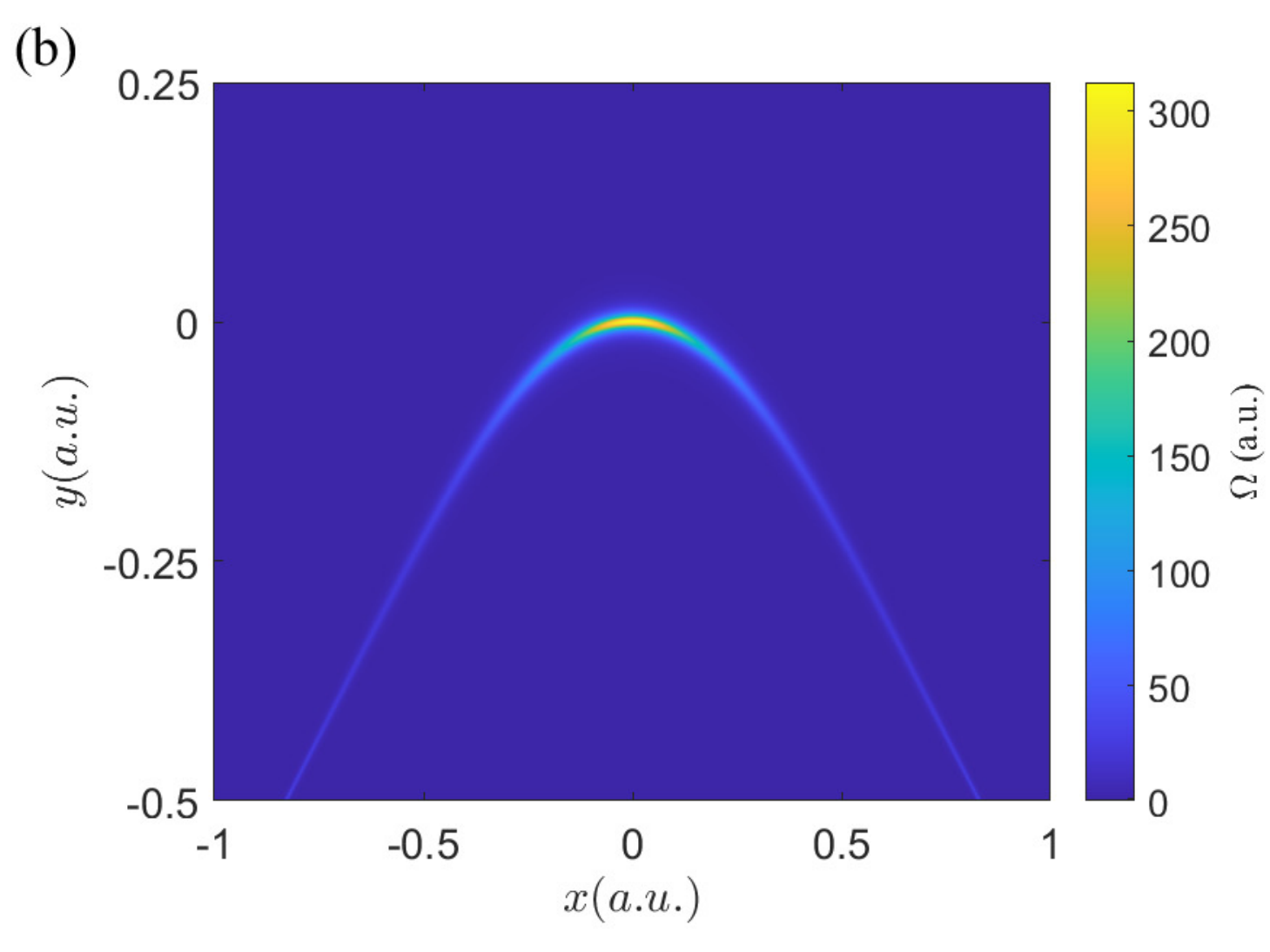}\label{fig:b}}
    \end{center}
    \caption{(a) The adiabatic potential surfaces for the model Hamiltonian (Eq.~\eqref{eq:h}).
    On the ground state, the incoming channel (marked by the magenta arrow) is located in $-y$ direction at $x=0$, the two outgoing channels (L and R, marked by the red and blue arrow according to their spin preferences) are located in $+y$ direction at $x=\pm r_0$. Nuclei wave packets with spin up electrons prefer channel R while nuclei with spin down electrons prefer channel L. The perturbed CI is at $x=0,y=0$. (b) The Berry curvature $\Omega$ of the ground adiabatic surface around the intersection point. Parameters are: $\mu=10^{-3}$, $\lambda=2\times10^{-4}$, $W=0.2$.}
\end{figure}

In this Letter, we will work with the following electronic Hamiltonian for the case of spin up electrons, with two electronic states (both with the same spin) and with two nuclear degrees of freedom, for which there is one incoming channel and two outgoing channels:
\begin{align} \label{eq:h}
    H = \begin{bmatrix}E_1(x,y) & V(x,y) \\ V(x,y)^* & E_2(x,y) \end{bmatrix}
\end{align}
where
\begin{align}
    &E_1(x,y) = A(e^{\epsilon_1 y} - 1) +\frac{1}{2}M\omega^2x^2 \nonumber \\
    &E_2(x,y) = \begin{cases} \frac{1}{2}M\omega^2{(\sqrt{(y-r_0)^2+x^2}-r_0)}^2,& y < r_0 \nonumber \\ 
    \min{\{\frac{1}{2}M\omega^2(x-r_0)^2,\frac{1}{2}M\omega^2(x+r_0)^2\}},& y \ge r_0 \end{cases} \nonumber \\
    &V(x,y) = (\mu x + i\lambda)e^{-\epsilon_2^2 y^2}
\end{align}
where $x,y$ are the nuclear coordinates. 
In our simulations, we set $A=0.02$, $\omega=0.01$, $M=10^3$, $\epsilon_1=2.5$, $\epsilon_2=2.5$, $r_0=2$ (all in atomic units).
In Fig.~\ref{fig:pe} we plot the adiabatic surfaces of $H$.
The real part of $V(x,y)$ (which equals $\mu x e^{-\epsilon_2^2 y^2}$) represents the diabatic coupling and the imaginary part $V(x,y)$ (which equals $\lambda e^{-\epsilon_2^2 y^2}$) represents the SOC.
Note that, if we ignore the SOC component of the Hamiltonian, there is a CI at $(0,0)$, just before the bifurcation of the two channels. 
However, the CI is removed when we add in the SOC, which should be a common situation for molecules with spin.
Note that, for this Hamiltonian, the phase variation parameter is $W = \pdv{x} \arctan{(\frac{\lambda}{\mu x})} = -\frac{\mu\lambda}{\mu^2x^2 + \lambda^2}$, which can be largest when $x \ll \frac{\lambda}{\mu}$, indicating that there could be a strong field effect in the vicinity of the origin.

Below, we will imagine a situation where a nuclear wavepacket approaches the avoided crossing from the $y \rightarrow -\infty$ channel
and then can emerge in one of two $y \to +\infty$ channels that are displaced in the $x$-direction:
The left (L) channel flows along $x = r_0$; the right (R) channel flows along $x = -r_0$. See Fig.~\ref{fig:pe}. 
For the Hamiltonian in Eq.~\eqref{eq:h}, we will show that the wavepacket chooses one channel predominantly over the other. 
Physically, this choice of channel means that a nuclear wavepacket with one spin (say spin up) will emerge into one channel. 
Now, recall that, if $H$ is the Hamiltonian of spin up electrons, then $H^*$ the Hamiltonian for spin down electrons because of time-reversibility.
Thus, because the Hamiltonian in Eq.~\eqref{eq:h} has the symmetry $H^*(x,y)=\sigma_z H(-x,y)\sigma_z$, it follows that, if incoming wave packets with one spin have a preference to emerge in one channel (say the channel at $x = +r_0$), incoming wave packets with other spin will have the exact same preference for the other channel (i.e. the channel at $x = -r_0$).
These spin-dependent nuclear scattering amplitudes are plotted heuristically in Fig. \ref{fig:pe}.


In Fig.~\ref{fig:b}, we plot the Berry curvature of the ground adiabatic state $\Omega = -i\bm{d}_{01}\times\bm{d}_{10}$. By definition, the Berry force that nuclei feel along the ground state
is $[{F_B}_x, {F_B}_y] = \hbar\Omega/M[p_y, -p_x]$. The Berry force of the ground-state surface is significant in a small region around the region of the avoided CI. Therefore, one might 
indeed expect that nuclei will experience a strong force when passing through such an ``avoided'' complex-valued CI region and there will be a large difference between the transmission of the outgoing terminals. 

\section{Results}



We have run scattering calculation for the Hamiltonian in Eq.~\eqref{eq:h} using the exact procedure outlined in Ref. \cite{wu2020chemical}. 
We calculate the transmission rates $T_L$ and $T_R$ for each channel in Fig.~\ref{fig:pe}. The nuclei enter asymptotically from from the $-y$ channel (with spin up and nuclear motion bound to the ground state in the $x$ direction); the nuclei can emerge in either the L or R channels.
The total incoming energy $E$ is defined by $E=p_y^2/2M + E_{\text{bound}} = p_y^2/2M + \hbar\omega/2 - A$.

\begin{figure}[H]
    \begin{center}
        \subfloat{\includegraphics[width=0.55\columnwidth]{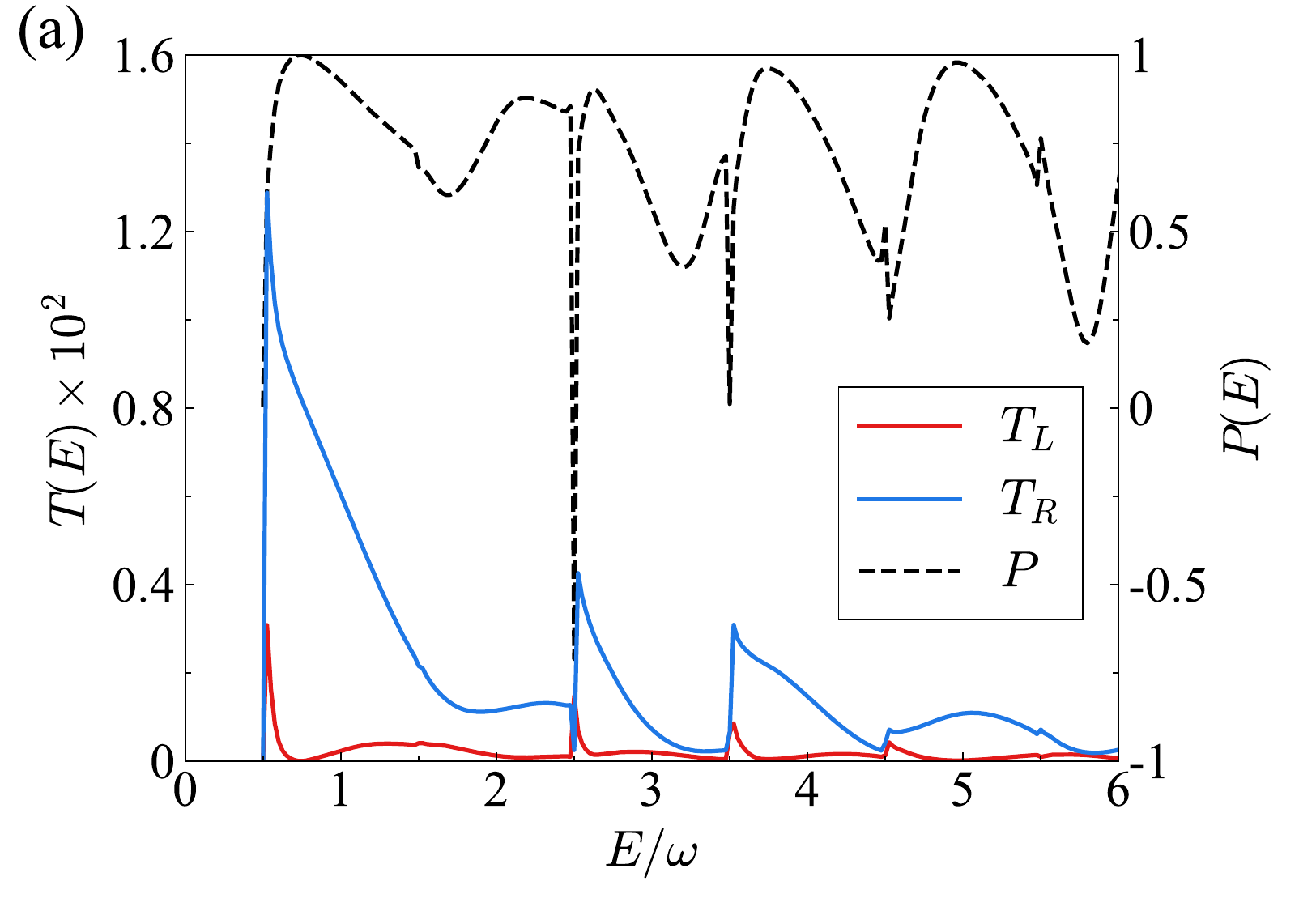} \label{fig:trans}}
        
        \subfloat{\includegraphics[width=0.5\columnwidth]{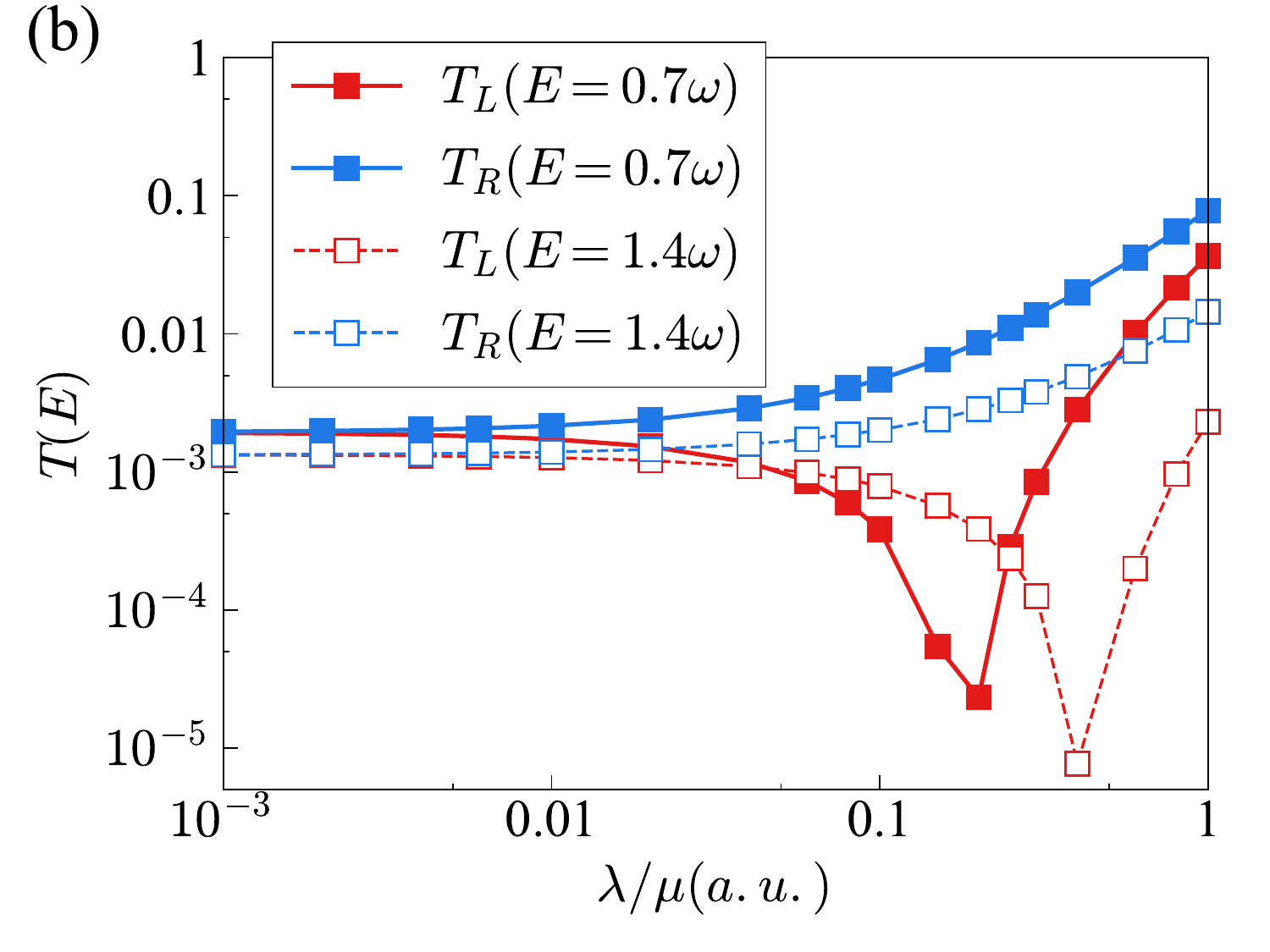} \label{fig:transw}}
        
        \subfloat{\includegraphics[width=0.5\columnwidth]{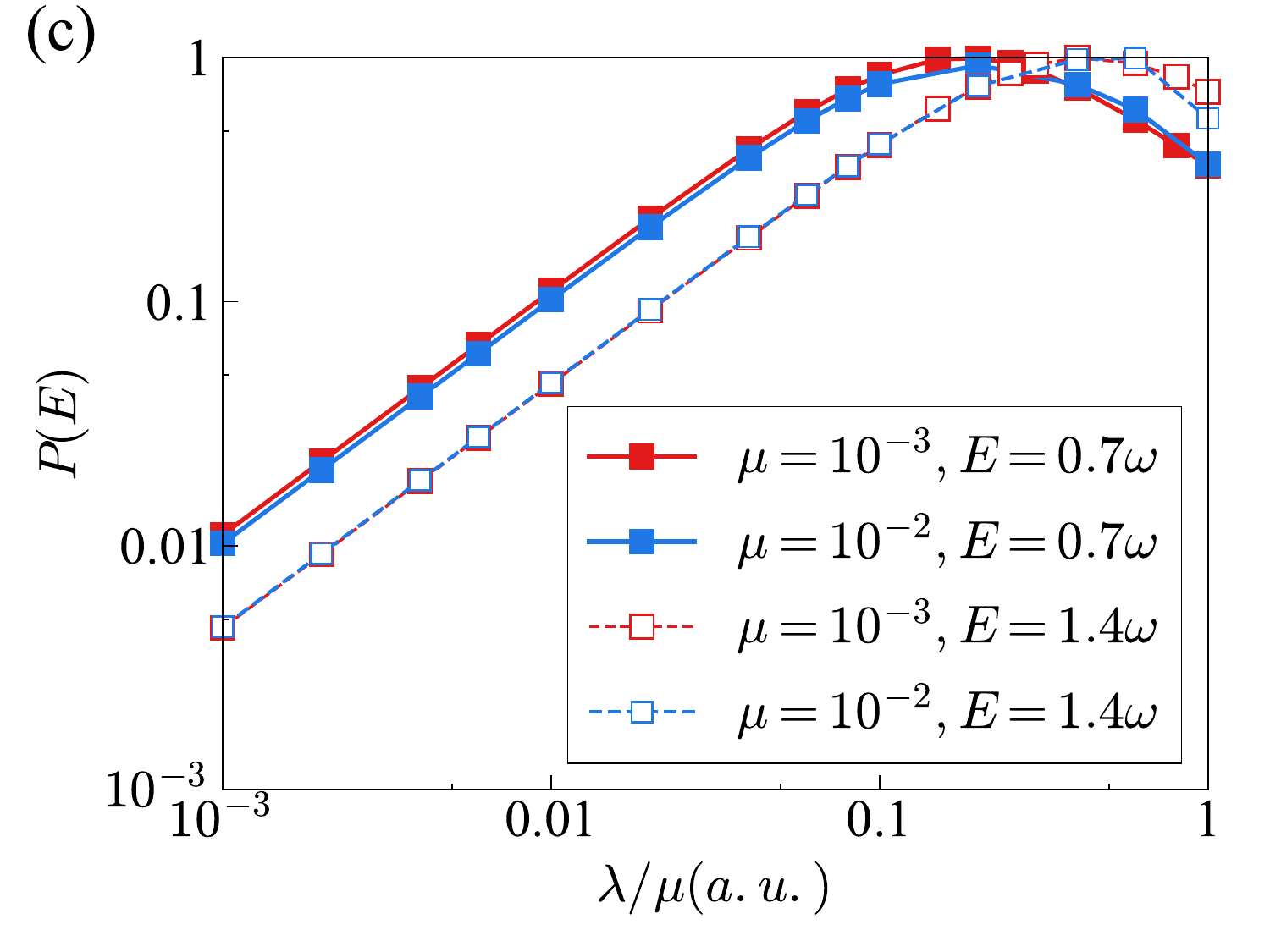} \label{fig:peak}}
    \end{center}

    \caption{(a) The transmission rates $T_L$,$T_R$, and selectivity $P=(T_R-T_L)/(T_R+T_L)$ as functions of  incoming energy for dynamics along the Hamiltonian in Eq.~\eqref{eq:h}. Note that the selectivity is very large, almost always larger than 50\% and often close to 100\%. (b) $T_L$,$T_R$ as function of $\lambda/\mu$ with two fixed incoming energies $E=0.7\omega$ and $1.4\omega$ and $\mu=10^{-3}$. While an increase in SOC ($\lambda$) always leads to an increase in $T_R$, this monotonic behavior is not true for $T_L$, leading to huge selectivity. (c) $P$ as a function of $\lambda/\mu$ at a incoming energies $E=0.7\omega$, $1.4\omega$ and $\mu=10^{-3}$, $10^{-2}$. Here, $P$ has little dependence on $\mu$ alone, but depends on the reduced parameter $\lambda/\mu$.}
    \label{fig:2}
\end{figure}

In Fig.~\ref{fig:trans}, we plot the individual transmission rates $T$ for the two channels $T_L$,$T_R$ as well as the spin selectivity $P\equiv(T_R-T_L)/(T_R+T_L)$, both as functions of the total incoming energy $E$. Here, one sees a huge preference for the right channel over almost all of the entire energy range. At certain incoming energies such as $E=0.7\omega$, $3.8\omega$ and $5.0\omega$, the selectivity is close to 1, such that nuclei with opposite electronic spins will be completely separated into the two outgoing channels.
Note that, due to the quantized nature of the transverse bound states, one find peaks and valleys in the total transmission as a function of the incoming energy. These oscillations in transmissions can also lead to spikes in the polarization, because the peak of $T_L$ often occurs at a slightly lower energy than the peak of $T_R$.
Nevertheless, overall, this figure highlights the fact that an ``avoided'' complex-valued CI can produce enormous spin selection.

Next, in Fig.~\ref{fig:transw}, we plot $T_L$, $T_R$ as a function of the reduced parameter $\lambda/\mu$ with a fixed $\mu=10^{-3}$ at two incoming energies $E=0.7\omega$ and $1.4\omega$. We find that $T_R$ increases monotonically with $\lambda/\mu$ for most values of $\lambda$, but  $T_L$ actually decreases with $\lambda/\mu$ until a minimum is reached. For a given total energy E, the maximum spin-selectivity is achieved when $T_L$ is minimized.

Finally, in Fig.~\ref{fig:peak}, we plot the polarization $P$ as a function of $\lambda/\mu$ with $\mu=10^{-3},10^{-2}$, again with two incoming energies ($E=0.7\omega$ and $1.4\omega$). Here, we see that  polarization changes very little with $\mu$ and depends (at least effectively) only on the reduced parameter $\lambda/\mu$. 
As stated above, $P$ has a maximum when $T_L$ is minimized, and this maximum is basically unity (100\% spin selectivity). We note that, even when $\lambda/\mu = 0.01$, one can find a 10\% preference for the right channel, which implies that even with a very small SOC, the presence of a Berry force around a complex-valued ``avoided'' conical intersection can lead to significant spin effects. 

\section{Discussion}

It is now fully appreciated that a huge number of photochemical processes in solution are indeed mediated through CIs; even though the seams of conical intersections arise with co-dimension 2 in configuration space, the effects of the enormous derivative couplings around CIs fill up configuration space and act as a funnel to bridge different electronic surfaces. In the present Letter, we have shown that, whenever spin-orbit coupling is present, these CIs can also potentially lead to something not well appreciated: non-trivial spin polarization as mediated by a Berry force.
As mentioned previously, Berry himself predicted that the Berry force associated with a true, unavoided complex-valued CI should {\em not} have an enormous effect on nonadiabatic dynamics; in practice, however, such a true complex-valued CI should also be exceedingly rare, and so this particular prediction is likely not relevant to photochemistry or chemical dynamics.

Looking forward, this Letter opens up many avenues for future experiments and theoretical investigations.
First, there are many examples of photoinduced spin chemistry in the literature, for which time-dependent EPR spectroscopy can measure unpaired electrons interacting with their environment \cite{hoff2012advanced}. Furthermore, there are also hosts of magnetic field effects within organic photochemistry \cite{Gould1984,Steiner1989,Hore2016,Hore2020} which have not yet be fully understood and which must eventually tie into the hot topic nowadays of avian magnetoreception \cite{Hore2016,Mouritsen2018}.
Within the current literature on organic photochemistry and spin-dependent chemical reactions, the impact of Berry forces has not yet been explored, which represents a huge opportunity for theoretical discovery.

Second, recent studies by Naaman {\em et al.} have shown that, even without photoexcitation or magnetic field effects, spin-dependent conductivity can arise when current is passed through chiral molecules, an effect known as chirality-induced-spin-selectivity (CISS) \cite{Gohler2011,Naaman2012,Naaman2015,Naaman2019,Naaman2020}.
To date, theory has been unable to explain why the CISS effect is as large as it is, given how small the SOC matrix elements are \cite{Naaman2015,Varela2016,Maslyuk2018,Naaman2019,Zollner2020,zollner2020a,Evers2020}. The present Letter must make one wonder whether conical intersections can be found within the manifold of conducting electronic states, such that a Berry force can help explain the large magnitude of spin selection.
We note that recent studies have shown that electron transfer in organic molecules such as DNA and proteins are largely incoherent \cite{Zhang2014,Xiang2015,Kim2016,Beratan2019}, raising the possibility that nuclear motion (and the associated Berry force) may well lead to spin selectivity. Moreover, the CISS effect has already been shown to lead to changes in over potential for water splitting and novel magnetic field effects \cite{Zhang2018}, suggesting that if we can indeed use Berry force to produce spin-selected molecular fragments, there may be a host of future applications including new spin-dependent catalytic mechanisms, exotic stable molecular spin devices, and efficient electrochemical metal-ion separation protocol.

Now, the experiments above represent interesting potential applications in spin chemistry and physics. At the same time, however, we must emphasize that, in order for practical progress to be made with quantitative experimental predictions, two theoretical questions will need to be addressed.
First, in the present Letter, we have used exact quantum mechanics to calculated scattering rates. These calculations are very expensive and do not always offer a simple explanation of the physics we observe.
More generally, for large systems, we will require new semiclassical tools (that treat nuclei classically) that are both inexpensive and that can offer intuitive pictures of electronic and spin relaxation.
Developing such tools (e.g. extending Tully's surface hopping algorithm \cite{Tully1990,Coker1995} to the case of complex-valued Hamiltonians) will be essential if we are to study systems with many electronic states and many nuclear degrees of freedom (ideally {\em ab initio} systems). 


Second, the question of exactly when and how Berry force and/or magnetic fields lead to observable effects in the condensed phase remains a general problem for spin chemistry.  
On the one hand, from a classical perspective, a magnetic field does not affect the equilibrium solution to a Fokker-Planck equation, and the magnitude of molecular SOC or hyperfine interactions are orders of magnitude smaller than the thermal energy $k_B T$ in the room temperature \cite{Steiner1989,Naaman2019}. Thus, one might be led to believe that magnetic field effects (and thus Berry force) must vanish with enough friction.
On the other hand, however, Berry force effects near a complex-valued avoided CI can be large.
Furthermore, for a molecule that is exposed to an out-of-equilibrium environment (e.g. a current runs through the molecule), the Fokker-Planck equation need not hold and there is no reason to expect that external friction will eliminate the influence of Berry force and/or spin polarization. For example, for a molecule near a metal surface, it will feel a Berry force in the form of an asymmetric electronic friction tensor when there is an electric current \cite{lu2012current,Bode2012,thomas2012scattering,dzhioev2013out,Dou2018,Dou2018a}.
Note that, Naaman and other researchers have shown that the CISS effect increases with increasing voltage such that the molecular dynamics is far from equilibrium \cite{Xie2011,Kettner2015,Kiran2016,Kettner2018,Naaman2020}.
Obviously, extrapolating from the present simulations to the condensed phase, and calculating the effect of a Berry-force induced spin polarization in the presence of friction, will be a crucial next step forward.
Clearly, several key obstacles remain if we are to ever merge theoretical chemistry with the field of spintronics.

\section{Conclusions}

In summary, we have investigated nonadiabatic dynamics for a model Hamiltonian $H = H_0 + H_{\text{SOC}}$ in the vicinity of a complex-valued avoided conical intersection (i.e. near geometries whereby $H_0$ alone would admit a real-valued conical intersection).
This model Hamiltonian exhibits a very strong spin selectivity of reaction pathways (close to 100\%) that remains very significant even when the SOC matrix elements are weak.
This Letter suggests that, in the future, simulations of nonadiabatic dynamics through conical intersections may find enormous spin polarization effects if SOC is included and Berry force is taken into account.
Furthermore, in practice, this Letter also highlights the possibility that, with a proper understanding of photochemical mechanisms, organic chemists may be able to synthesize molecules that ensure spin-selection, thus taking a very different approach towards the development of spintronics.

\begin{acknowledgments}
This work was funded by the Center for Sustainable Separation of Metals, an NSF Center for Chemical Innovation (CCI) Grant CHE-1925708.
\end{acknowledgments}

\bibliography{complexJ} 
\bibliographystyle{apsrev}

\end{document}